# A NOVEL DATATYPE ARCHITECTURE SUPPORT FOR PROGRAMMING LANGUAGES

Mehran Alidoost Nia and Reza Ebrahimi Atani

Department of Computer Engineering, University of Guilan, P.O. Box 3756, Rasht, Iran.
alidoost@msc.guilan.ac.ir, rebrahimi@guilan.ac.ir

*ABSTRACT*

*In programmers point of view, Datatypes in programming language level have a simple description but inside hardware, huge machine codes are responsible to describe type features. Datatype architecture design is a novel approach to match programming features along with hardware design. In this paper a novel Data type-Based Code Reducer (TYPELINE) architecture is proposed and implemented according to significant data types (SDT) of programming languages. TYPELINE uses TEUs for processing various SDT operations. This architecture design leads to reducing the number of machine codes, and increases execution speed, and also improves some parallelism level. This is because this architecture supports some operation for the execution of Abstract Data Types in parallel. Also it ensures to maintain data type features and entire application level specifications using the proposed type conversion unit. This framework includes compiler level identifying execution modes and memory management unit for decreasing object read/write in heap memory by ISA support. This energy-efficient architecture is completely compatible with object oriented programming languages and in combination mode it can process complex C++ data structures with respect to parallel TYPELINE architecture support.*

*KEYWORDS*

*Significant Data Type, Programing Language, dark Silicon, implementation, ISA,*

## 1. INTRODUCTION

In today's world, programing languages such as Java apply advance data types to increase the flexibility in software implementations. In the other hand, when it comes to hardware design there are several difficulties which does not let the designers to follow a fruit full implementation. It is hard to train computers, understand all aspects of programming languages. It comes from high level perspectives in programming tools and also is related to low level design in computer architecture. This misunderstanding between programming languages and computer architecture needs an interface for interpretation which leads to complicated design architecture. In order to solve this problem, we can use some specifications of programming languages which helps improving the hardware ability of understanding codes and a clear and easier execution.

Datatype-based architecture effects on some current issues in computer design. Energy efficient architecture design, reducing execution time, increasing overall speed, some parallelism level and also object oriented support are some aspects of type-based architecture improvements. Prior works in these areas come from different ideas. Conservation core, *Greendroid* project, dual voltage in Truffle architecture to support approximate programming and Green framework are previous works on energy efficient architecture design [1, 2, 3, 4]. Also ICER is implemented respect to Code reduction technique to reduce energy consumption [5, 6]. Object oriented architectures are focused on object management in heap memory and increase speed of memory allocation by using compaction algorithms to handle object memory update [7, 8]. Compiler support is needed to carry out instruction level parallelism and static reordering. For instance Enerj compiler is provided to determine approximate-aware regions and LLVM compiler that completely support object oriented programming optimization [9, 10]. Parallelism tools in code level support efficient implementation of parallel execution and include some specifications of code regions that can help the term of parallelism. We should know what region of codes are more deserved to support. Parkour and Kremlin are previous works on this area [11, 12].

Datatypes have significant advantages that make us moving toward datatype-based architecture design. One of the most important properties is to reduce type annotation. Before this, an approach called TAL was used for providing security in programming language level. Type assembly maps all different datatypes into a unique datatype template which is defined for security reasons. It wasted compiling time in several steps hence iTALX is derived from TAL which reduces a large number of code caused by type annotations [13, 14]. Mentioned effects are useful in memory management design and release occupied dynamic memory. Also store operations and memory management can be improved by using datatype architectures. ADI instruction that is defined according to datatype and is tested under oracle and some other storing environment is led to reduce energy consumption [15]. In object oriented programming languages like C++, it is efficient to build specific object file. This can help us to improve performance of the execution code and increases ability to adapt with new architecture design. ELF-like object file is completely sufficient [16, 17].

In this paper we focus on data type features in programming languages that is needed to support through architecture design. In practice, most of datatypes are common among languages due to code behaviour. Dividing design space into specific regions relevant to datatype features leads to increase process speed, improve parallelism and also reduce overall energy consumption in some cases. Each datatype has its own privacy and data management and this is equivalent to object concepts in object oriented programming and implementation of advanced programming languages. But still a challenge is remained. What datatypes should be chosen to implement through TYPLINE architecture? We answer to this essential question after type analysis process. However application specific platforms with their homogenous datatypes are more compatible, but in this paper we are trying to obtain TYPLINE performance on general benchmarks in parallelism, speedup and also energy-efficiency. This helps us to understand real targets of this architecture and also identify hot regions to implement.

The rest of the paper is structured as below: In section two a brief overview of datatype architecture is presented. Datatypes in different applications are analysed in section three and in section four TYPLINE, a novel datatype-based architecture is introduced. The evaluation results of the architecture are presented in section five and we will have a comparison with other related architectures in section six. Finally the paper concludes in section seven.

Data types are the best target to cluster operations into specific regions. Boundaries between these regions should be determined by type of operations. The leverage of Datatypes comes from programming languages approaches to identify operation types. Datatype architecture idea is behind programming languages behaviour against types of data and also types of operations. So it is strongly connected to programing language scope. Conceptions like heap memory, objects, linker, static and dynamic type checking, method signatures and related topics should be considered in architecture design.

As shown in Figure1, we faced to equivalent conceptions in programming and hardware level. To completely support language level abstractions, it is needed to know programming language scopes and its level of cooperation among other units like operating systems and main hardware units.

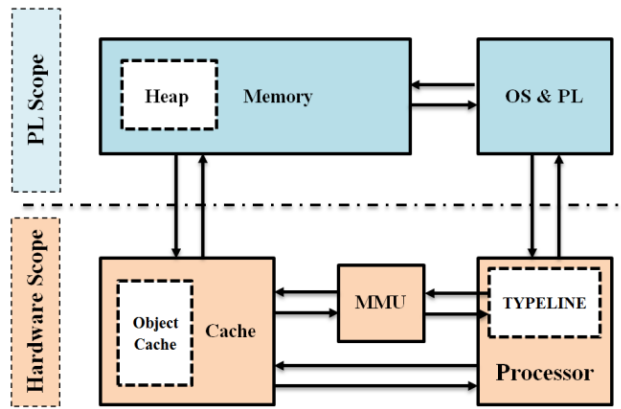

Figure1: Programming language scope versus hardware scope is shown in this figure. Heap memory in programming language scope should be mapped into object cache in TYPELINE mode in hardware level.

## 1.1. Design Steps

Datatype architecture is consisting of some process lines that in special situations can be executed as parallel. It is responsible for running typical datatype operation is a processor unit. The main idea of this architecture design is to map language level feature in hardware level to reduce size of codes and also annotation needed for description and support abstraction data types (ADT) to run simultaneously as much as possible. In this architecture each datatype is executed in its own process line and operation support is related to the same datatype. However according to current situations, it is not possible to dedicate design space to all datatypes in parallel, but it will be shown that correct election of target datatypes can be improved process speed and parallelism levels. It is important how to divide design space. It must be cleared that how many silicon space should be dedicated to specific datatype. For example if floating point operations have more frequency in comparison with integer, it is clear that we should dedicate more design resources to that.

First step is to find most important datatypes with respect to operations that should be executed. This step is explained in the next section that needed statistical analysis in general benchmarks. After identifying target datatypes, it's time to specify design regions. In this paper we are trying to use four most important datatypes that can be run and work together simultaneously. In paralleling design space, one of the problems is power consumption. It seems very critical to handle all paralleled stages with the same acceptable power. For example if we have a power source with 80W, overall power requirement to trigger all parallel units must not exceed 80W. As mentioned, each process line can work independently or parallel in cooperation with other lines. So it is needed to propose an instruction set (ISA) for architecture support. Each line has its own data and logical operation instruction and in some special modes, process lines can be reconfigured with respect to the current execution mode. Also this framework requires compiler static support to determine which code regions should be executed in which process line and absolutely identifies current execution mode. One of basic features of this instruction set is memory management instructions. It is common among all process lines and is focused on object file management that should be new and delete along with program running. This instructions can help to reduce amount of memory space usage during program execution. In addition, it can be considered as architecture support for object oriented languages that enhances entire performance in memory and cache level.

Next, we need datatype features and operation needed that to be implemented. In architecture design space, we should emphasis on language features. Type behavior in every typical language is very near but for better analysis, in this paper is focused on C++ language feature and all descriptions are completely compatible with this.

## 1.2. Type Measurement

Another concept that should be explained is to introduce a measurement for determining type of operations. This measurement is related to method signatures in language level. During type checking process in compiler level, type consistency is checked and type conversion that are performed. The here, all inputs and outputs for operations, methods, objects and other executive units are determined. So it should be decided that current operation must be executed in which process line. Method signatures are the main resource for specifying true process line. Assume that we have a heterogonous type operation including char, integer and floating point. After type conversion process, final method signature for entire operation chooses as float. Although the correct answer is clear with respect to C++ standard compilers, but as a standard rule, method signatures should be responsible for determining type of operations.

$$Float = float * char * integer \text{ (star is a sample operation)}$$

## 1.3. Process Lines and TEU

Process lines are the main part of datatype architectures. After detecting type of operations in static code analysis, it's time to direct them to the correct process line. Each process line leads to specific datatype unit called TEU (Type Execution Unit). Here can be performed single operations or parallel operations in cooperation with other process lines. Parallel mode should be determined in compilation time and cluster these operations into first logical unit.

As shown in Figure 2, in datatype architecture design we are trying to decompose entire operations into clusters that can be supported by TEUs. After statistical analysis and design steps, datatypes including B, D, A and G are chosen as the most significant datatypes. Green circles indicate these datatypes. Blue circles is some ADTs, structures, methods and functions that are constructed of the significant datatypes. These datatypes are the main target for parallel execution. In Figure 2-b the parallelism is shown by connected lines between two or three process line that ended to the TEUs. Other remaining datatypes could be handled in two ways. First is to use traditional processor as a secondary process line to handle other datatypes. Second is to convert all datatypes into significant datatypes that TEUs are enabling to run them. This conversion can be done by Type Conversion Unit that is discussed in section four.

Resources are assigned to process with respect to type value. In the other word, process lines that are responsible for frequent type operations need a lot of design space and also more computer resources. Datatype B with blue colour is the most significant in comparison with others. Consequently it is deserved to have more design space and first class priority in running. In this figure, when we move from first process line to the last one, congestion of related operations in process lines will be decreased. This is the direct result of similar operations and choosing correct implementation.

Assume that an ADT is constructed of three important datatypes including B, D and A. when datatypes in an ADT are heterogonous and behave uniformly, we can propagate execution through available process lines that are provided here. So according to Figure 2, in the first layer near TEUs, three datatype B, D and A are connected together by vertical line. This is because in logical state these three separated datatypes comprise single ADT and in this architecture can be executed simultaneously.

This shows the parallelism feature in datatype architecture design. If the number of these features enhance, performance will be increased.

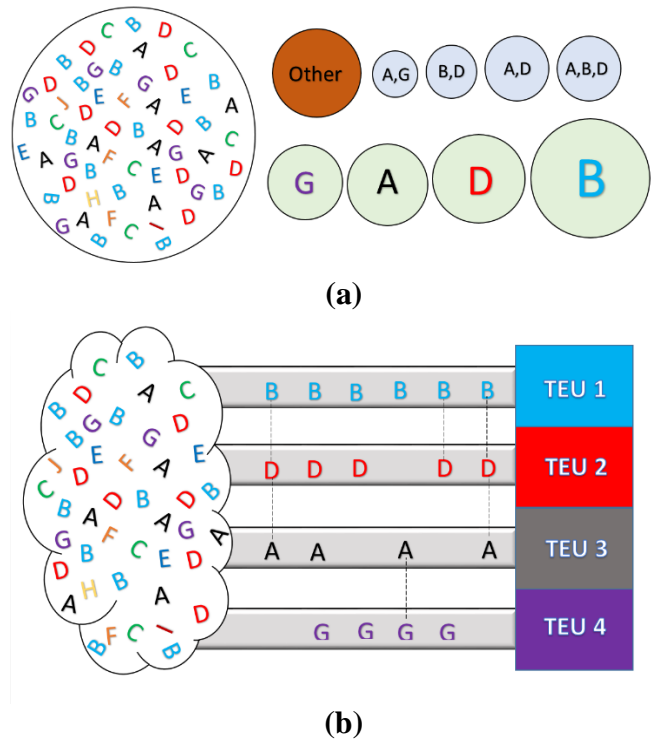

Figure 2: Datatype architecture design steps. (a) It shows the first step including cluster of current datatypes in execution code. This step is static datatype ordering. (b) It indicates operation execution handling according to datatype support and also shows complex dynamic datatype decomposition simultaneously. (English alphabets are equivalent to datatypes)

### 1.4. Datatypes and Code Dependency

One of the most important issues in parallelism is the code dependency. It is behind dependency between related codes with common variables. In view of datatypes, the only measurement to separate operations and make a boundary, is the type of operations. When we are moving toward datatype approaches, it is more important to know about datatype behavior and structure in various situations. Data structure elements are more independent than others. It comes from data records that are define in a specific data structure and often are declared to show independent elements of a record. For example it is very rarely to define x variant in data structure and also define $x^2$. Even this happens, this variants are initialized together.

This theory is a strong reason to use datatype as a basic measurement to category process lines. Generally using datatype features to cluster process lines causes that code dependency is decreased sensitively and also it reinforces instruction level parallelism. A datatype is responsible to manage its own data and variable values and it is completely equivalent to object theory [18]. Objects have their own private data and this provides object oriented support availability. These concepts are discussed in section six.

### 2. SDT Selection

It is important to know that how to divide design space into useful regions. When applications become more heterogonous, the election of true design space will be more difficult [19]. These regions are given by statistical analysis, especially on hot code regions. In this section we are going to see experimental result behind codes. Code level includes standard C++ benchmark which is shown in Table1. We have tried to collect features related to implementation steps and compilation time which are important to determine exact hardware design. For this reason some benchmarks were selected through useful and common applications in different aspects of implementation including various arithmetic and logic operations that have been used by popular datatypes.

### 2.1. Datatype Analysis

Our goal is to find Significant Datatypes (SDT) through common applications. This step is required to know more details about implementation features of data types and statistical results of these features during execution time and code structure in native code side. Then we will show that significant data types based on some specific features related to programming languages features have a limited members. It means that SDTs in various implementations will be constant and stable than overall application features. After specifying SDTs, it is needed to identify special effect of data types on architecture design. We are going to approve that using SDTs in design steps which leads to increase the overall execution performance especially when the number of loop counter grows fast. For example in operating systems that one or more specific routines will be executed several times in an endless loop, datatype architecture is the best solution. But it should be noticed that in various platforms selecting SDTs will be different. For example in Windows OS we are faced to integer family like long, short and etc. Features like this leads to design interactive architectures that can be worked across programming language cooperation.

Table 1: Standard Benchmark for Type Analysis

| Title | Description |
|---|---|
| **Astar** | Path finding algorithm |
| **Btree** | Binary tree indexing routines |
| **Fss** | Fast SAT Solver |
| **Fft** | Fast Fourier Transform |
| **Viterbi** | The basic Viterbi decoder operation, called a "butterfly" |
| **Raytracer** | 3D image renderer |

Selecting best SDTs depends on two important factors. As shown in Figure 3(a), number of datatype repetition and according to Figure 3(c), number of program loops will be crucial in SDT choosing steps. Number of repetitions should be most important because they consist of large number of datatypes in program execution. It shows in Figure 3(b) as avg columns.

| Benchmark | int | float | double | long | char | struct | enum | typedef |
|---|---|---|---|---|---|---|---|---|
| astar | ↗ 61 | ↓ 0 | ↓ 0 | ↓ 0 | ↓ 1 | ↓ 0 | ↓ 0 | ↓ 0 |
| btree | ↓ 21 | ↓ 0 | ↓ 0 | ↓ 4 | ↘ 41 | ↓ 2 | ↓ 0 | ↓ 5 |
| fss | ⇑ 144 | ↘ 38 | ↓ 7 | ↓ 16 | ↓ 6 | ↘ 47 | ↓ 4 | ↓ 12 |
| fft | ↘ 31 | ↓ 0 | ↘ 36 | ↓ 0 | ↓ 0 | ↓ 0 | ↓ 0 | ↓ 0 |
| viterbi | ↓ 7 | ↓ 0 | ↓ 0 | ↓ 17 | ↓ 0 | ↓ 0 | ↓ 0 | ↓ 1 |
| raytracer | ⇑ 221 | ⇑ 151 | ↘ 37 | ↓ 0 | ↘ 41 | ↓ 23 | ↓ 10 | ↓ 4 |

(a)

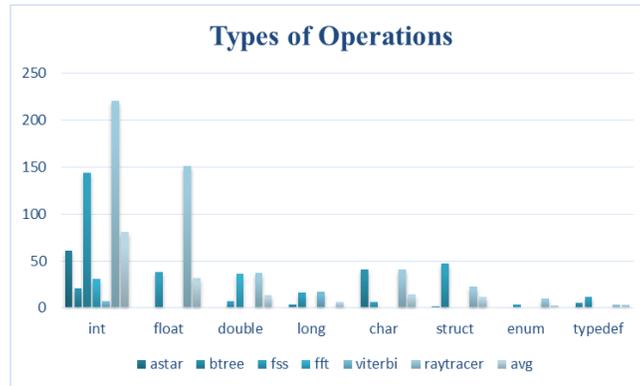

(b)

| benchmark | loops | condition | static | const | unsigned | array operation |
|---|---|---|---|---|---|---|
| astar | ↓ 25 | ⇑ 95 | ↓ 0 | ↓ 4 | ↓ 0 | ⇑ 112 |
| btree | ↓ 23 | ↗ 63 | ↓ 3 | ↓ 21 | ↘ 34 | ↘ 35 |
| fss | ↘ 37 | ⇑ 98 | ↓ 21 | ⇑ 107 | ↓ 4 | ↓ 15 |
| fft | ↓ 9 | ↓ 7 | ↓ 10 | ⇑ 117 | ↓ 28 | ↓ 6 |
| viterbi | ↓ 7 | ↓ 13 | ↓ 0 | ↓ 2 | ↓ 1 | ↓ 21 |
| raytracer | ↘ 40 | ⇑ 102 | ↘ 40 | ⇑ 155 | ↘ 32 | ↗ 68 |

(c)

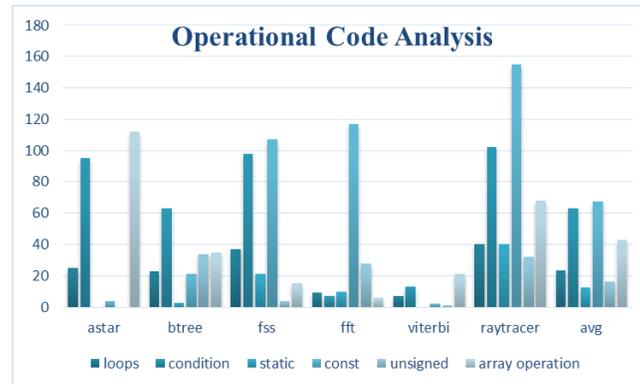

(d)

Figure 3: Operational datatype analysis. (a, b) Show datatype operation frequencies in benchmarks. (c, d) Indicate operation features according to code behaviour including loops, conditions, static operations, constants, unsigned and array operations. (The value of the arrows is raised respectively by colour order red, yellow and green)

Average number of the entire data type repetitions represents importance level of a specific datatype. But in choosing steps we need more exact parameters. It is because hardware implementation of these datatypes leads to spend design cost of use. Then we must quietly careful in selecting SDT procedure. Hence we follow a complementary factor that can help to make a better decision. This factor is related to data type features. Figure 3(c, d) indicates statistics of data type features according to standard benchmarks.

## 2.2. Operational Type Features

When it comes to program codes, the first and most important feature in program codes is loop repetitions. Datatypes have a fixed number in each program. But we should attend to datatype operations through loops. Loops will be multiply data type power in programming languages. Consider a program with repetition of 1000 times in a loop with 2 data types and another program without any loop with 100 of the same data type. It is clear that first program is more powerful in datatype level. Datatypes that are repeated in loop environment are our target for SDT. Priority of data type selection for hardware implementation is related to the same data type features. Notice that green arrows in Figure 3(c) show power of targets in comparison with yellow and red arrows.

Other features like conditions, tell us to focus on logical operations in architecture design. Static feature indicates values that should be more stable in heap and cache memory during execution time. Method signatures that have specific input datatypes and also have a specific output datatype that have a static feature are targets for object cache modelling. Also *const* and *unsigned* features can be implemented in hardware level as a feature and reduces number of type annotations according to their frequencies.

## 2.3. Array Operations

As shown in Figure 3, number of array data operations is impressive and also according to benchmark codes, most of them are in loop chains. After brief review in statistics that are shown in figure 3(c, d), it is clear that we should design the entire architecture with respect to array datatype support. It is because of the number of array operations in datatype features. Large number of operations is forced us to choose arrays as a SDT feature.

As a conclusion of this section, some of these datatypes is selected as SDT. These SDTs include integer, float, double and char. This selection is respect to type analysis and operational features that completely described in this section. It should be remembered that we are used general benchmarks, hence the result of datatypes is so vast. This approach is used for general purpose designs that should work in different situations. But choosing application specific targets are more useful and leads to receiving smooth results. This unification can increase performance very fast. Consider that we want to design a datatype architecture for DNS servers. DNS servers have a specific data template for exchange queries between users. This unification should be clearer where the special duty of target architecture is to compute and translate one million queries every day. When datatypes are in a unique template under DNS operations, it can handles more queries by process lines and also can be more focused on parallelism issues. Absolutely DNS datatype architecture design is more effective than general purpose architecture. So if execution codes are smoother, performance of the datatype architecture will be improved.

## 3. TYPELINE

Process line idea plus according to datatype theory leads to design a novel datatype architecture called TYPELINE. In this section we are going to explain TYPELINE features and its functionalities in detailed. Preliminaries and important issues in datatype architecture was described in two previous sections. So far, significant concepts in programming language scopes, process line idea, datatype theory, parallelism issues and high performance datatype architectures are described. It is indicated how to dedicate better design space according to datatypes. The procedure include detecting SDTs is mentioned and here we should use these important features that are extracted from operational code analysis steps.

Now we want to discuss about design steps in two important subsections. The first part is the implementation of TYPELINE architecture with respect to SDTs and code analysis and the second part is, memory operations and consequently issues related to object memory management. Through these steps, special ISA for this architecture design according to Alpha instruction set is proposed. This instruction set supports critical memory operations and designs with respect to each SDT features. Also control path of TYPELINE should be reviewed.

### 3.1. TYPELINE Architecture Design

A significant procedure is related to static code ordering. To handle this step, an appropriate compiler framework is needed to determine code regions that should be transferred to the correct process line. In this process execution mode is identified and C++ codes are translated according to proposed ISA. This step is completely static and codes have only limited dynamic availability. So process lines of the TYPELINE are equivalent to in-order pipeline and its behavior is similar to it. The dynamic availability is beyond configuration of some control states that can be changed execution mode and type features. As mentioned in previous sections, code dependencies in static analysis will be determined and parallel execution through various process lines is related to datatype of operations. With respect to type of operations, it will be really simple to configure process lines during execution time. Type conversion is also available in both static and dynamic state but it should be handled in static level. Dynamic datatype conversion is only because some of acceleration in execution time, should be applied that can improve overall speed of TYPELINE architecture.

As shown in Figure 4, TYPELINE consists of four process lines that can work independent or simultaneously. This number is because of four SDTs that extracted in type analysis step. Each process line begins with register files and ends to one TEU. Each register file includes 32 registers according to the same type features that can be loaded by control bits in two different modes. First is the single mode and another is array mode. In array mode, from 1 to 16 concurrent operation can be done by TEU.

After the operation accomplished, it is time to store results. Results can be write back to register files for future related operations or store in the memory. This transfer is done by global data path that responsible for data exchanges. The overall architecture is designed in a way that process lines will not be idle or sparse. We are tried to do best selections in type of operations.

TEUs are dedicated to four SDTs respectively integer, float, double and char. For design space issue division operation is skipped and it handles classically. Also in TEU 4 that is assigned to char operations, only sum, subtraction and logical operation is supported. This is because in char operations or string-related processes we are involving to comparison and logical operations instead of computation operations. The functionalities of each TEU unit is completely explained in ISA subsection.

Exceptional operations or rare functions that are not support in these four process lines can be handled in two ways. First is to convert them by type conversions and then run across process lines and second is to use a complementary classic computation units that supports traditional machine codes. In this project we are chosen the second solution. This selection is because first solution will be led to data lose in type conversion processes and also reduce overall operational performance of the entire architecture. So this is wisely to use traditional support as secondary process handling. This act is important where it is more significant to ensure process reliability. For any reason, if type processing is failed, it ensures that process will be accomplished.

Due to the constraints of power budget, it is reasonably to economize design space by choosing correct computation and logical units in TEUs. The main assumption about power source is that we have 80W power for providing entire energy consumption. According to dark silicon problem we faced to power constraints for thermal reasons [20]. It is clear that we have not wide choices. Consider 300 mm$^2$ design space under 90 nm TSMC technology. For using all transistors in this area, it is required approximately 445W [1]. But in this situations only about 18% of design space can be used with 80W power budget. So in parallel computational architecture it is very important to attend to these limitations. Another solution is to use low-power techniques to reduce overall energy consumption that are more common in synthesizable architectures and can be used during implementation steps [21]. Design optimization seems very necessary during implementation procedure. This feature can robust the entire performance of TYPELINE.

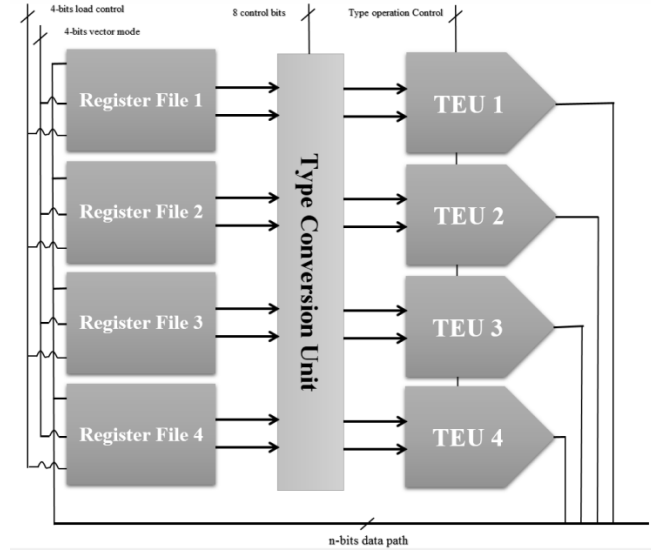

Figure 4: An overview of TYPELINE architecture

## 3.2. Instruction Set Architecture

As explained it is necessary to have special instructions for each of TEUs according to type of operations. So we are proposed instructions needed due to Alpha ISA [22]. The instructions are provided in Table 2. It refers to previous sections that we are mentioned conventional part of processing should be handled by traditional instructions under Alpha architecture.

As shown in Table 2, instructions are divided into 6 categories. The first four groups are related to operations that can support same datatype. Also it should be noticed that in popular datatypes like integer, more design space is dedicated and consequently the number of instruction support is higher than other datatypes. Although in integer datatype is emphasized on both arithmetic and logical operations but in double and float datatype is attended to arithmetic and finally char datatype operations are included comparison and logical instructions. These instructions are extracted due to datatype analysis in section three and are proposed to support datatype functionalities according to its operational behaviors.

It is time to explain control instructions. With respect to TYPELINE architecture, it is required to determine whether we want to use vector mode for parallel purposes. VEN (Vector Enabled) and VDS (Vector Disable) instruction is used to change between single and vector load in register files and also determine operations should be executed single or parallel in the same TEU. It configures these four control bits. Also it is possible to enable or disable process lines manually. Only integer datatype operations are protected. These availabilities are because in some situations we want to focus on single or double datatype operations or critical computations that are needed to execute reliably. The instruction include FTEN, DBEN, CHEN, FTDS, DBDS and CHDS are provided for supporting these features. CONV instruction is proposed to configure type conversion unit manually. It configures by 8 length binary string that each bit of it indicates one type conversion state. Type conversions can be applied to 8 convertor with respect to 8 control bits. But according to type constraints in TYPELINE architecture, only three type conversion switch is performed that are responsible for integer to float, integer to double and float to double conversion. From reserved switches can be used to enhance performance of processing. For example if we want to process a heavy array, it can be divided into other process line. It is possible only with type conversion unit.

PEN instruction puts execution control in parallel mode. Every codes that come after this instruction, can be executed simultaneously according to their process lines. This mode is used to support ADT and user defined datatypes and objects. After entire process is done, in should be disabled by PDS instruction.

Table 2: Proposed instruction set for TYPELINE.

| Category | Instructions |
|---|---|
| integer | LD.in, ST.in, MOV.in, ADD.in, SUB.in, MUL.in, DIV.in, CMPE.in, CMPEG.in, CMPES.in, CMPS.in, AND.in, OR.in, XOR.in, NOR.in, XNOR.in, SRA.in, SRL.in |
| float | LD.ft, ST.ft, MOV.ft, ADD.ft, SUB.ft, MUL.ft, DIV.ft, CMP.ft |
| double | LD.db, ST.db, MOV.db, ADD.db, SUB.db, MUL.db, CMP.db |
| char | LD.ch, ST.ch, MOV.ch, ADD.ch, SUB.ch, CMPE.ch, CMPEG.ch, CMPES.ch, CMPS.ch, AND.ch, OR.ch, XOR.ch, NOR.ch, XNOR.ch |
| Control | VEN, VDS, PEN, PDS, FTEN, DBEN, CHEN, FTDS, DBDS, CHDS, CONV |
| Memory | OBJ.n, OBJ.r |

Object oriented architectures have their own dynamic memory support especially object cache. OOMIPS is one of architectures that support object oriented dynamic memory management [7]. In that architecture allocation and releasing of the dynamic memory is done by messaging system. This message line is a direct communication among processor, object cache and main memory. But this communication system is not optimized because of message management overheads. For this reason, we are proposed object memory management operations that can do these duties. The OBJ.n instruction is responsible for allocating new object space in heap memory and OBJ.r removes data objects that have been used, from dynamic memory.

The main reason that we need object memory instructions is to reduce memory management time and save wasted memory that is occupied by useless object elements [7]. Using these object memory instructions in object oriented architecture can reduce garbage in dynamic memory. This problem is caused by programmers and in programming language steps that grant users the availabilities related to allocating dynamic memory. For instance, assume that in an endless loop we allocate heap memory to an object. But after that object's functionality is done, the object stays in heap memory until program is died. Often programmers only allocate the memory space and most of these situations, programming language is responsible for managing data objects in dynamic memory. In this framework we are trying to use object instructions to quickly support memory management.

### 3.3. Through the Codes

In this subsection we are going to review a simple example of code executing on TYPELINE. A very intelligent approach is clustering. It means that user defined and independent datatype operations can be executed simultaneously into one cluster. This idea comes from VMWare code translation to handle heavy exits between privileged and de-privileged instructions [23]. We use this idea with some difference in implementation. PEN instruction is determined cluster beginning and maximum capacity of a cluster should not exceed 4 datatype operations.

Figure 5 is shown a simple code execution steps through TYPELINE. The first green triangle represents compilation step. Vector mode is prepared for uniform datatype loading and is began from first integer load operation to the end of current cluster. The cluster size is chosen dynamic. But it is important to know that cluster members in memory operations should not exceed 16 (according to register constraints) and also in arithmetic and logical operation should not be up to 4.

As mentioned rules, we have two clusters in these codes. First is about load operations and second is arithmetic operation. Notice that in load clusters all members should be in the same category and in operation clusters, members should have completely different datatypes (like this example) or all members should have common datatype. In this code we have two cluster that can be ran simultaneously. But the result of execution is more significant. In first cluster with spending one LD.in operation clock cycles, we can load four registers simultaneously.

Also in second cluster we faced to parallel execution of ADD.in and DIV.ft. CONV instruction can be executed in 1 cycle and overall cycles for executing this cluster is equivalent to 1+DIV.ft. It is because ADD.in absolutely takes less cycles than DIV.ft and one cycle is needed to prepare configuration for type conversion unit. In addition, it is saved (3*LD.in+DIV.ft+1) cycles to run this code region. If the various datatype operations are distributed in a higher level among code regions or the number of specific datatype grows in code regions, the TYPELINE works better and idle process lines will be decreased. As idle lines decrease, the entire performance will be enhanced directly. Here, performance enhancement is related to overall speedup of the TYPELINE. This is the direct result of parallel execution and cycle reduction.

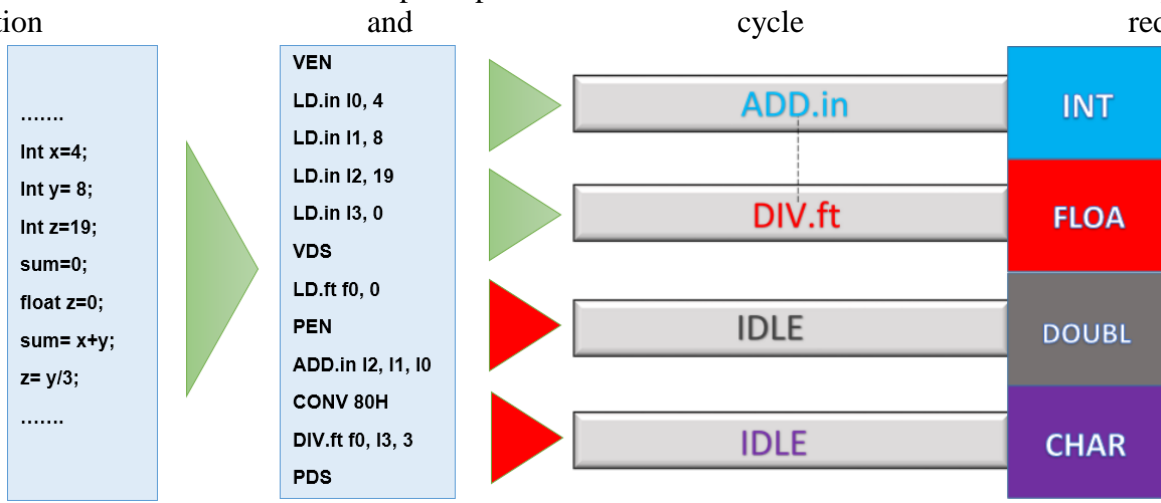

Figure 5: It shows code execution steps through a simple example. After translation and compilation steps in parallel mode, two of process lines get idle and only two real parallel execution exist. Red triangles represents idle and useless computation channels in this example.

## 4. Experimental Results

In this section we are going to review TYPELINE effects on three aspects of execution include load parallelism, computational parallelism and reducing code execution cycles through standard benchmarks and experimental statistics.

After the compilation process is completely done, it is time to prepare codes for execution. This preparation include some optimization steps that is related directly to programming language and also compiler that is used to cluster machine codes. The order of machine codes have maximum side effect on number of parallelism in load instructions. This is because we cluster code regions with respect to their order in original source code. If the order can be changed by compilers, final results may increase approximately to 5%. The main factor in computational instruction parallelism is to have uniform datatype operations in code regions or number of complex datatypes, user defines and structures increase. Using datatypes according to process lines support can enhance overall performance and decreases entire time that process lines get idle.

Figure 6 indicates approximate load parallelism, computation parallelism and cycle reduction in different benchmarks. As shown in (a) raytracer has the best parallelism availability in load instructions and the FSS is the bad one. Average parallelism in load instruction is about 40%. As mentioned, the parallelism in load instructions is directly related to the code orders and type of load instructions that come continuously. Due to the section three, raytracer has the best distribution in type of operations and FSS despite to have strong datatype operations but distribution of load instructions is very sparse and this is the main reason that its code regions is not compatible with parallelism concepts in TYPELINE architecture. So uniform load instructions is seen in application- specific environments more than others.

In (b) it is presented that computation parallelism is so lower than load instructions. In average, parallelism is about 18%. It means benchmark with various functionalities are not main target of TYPELINE architecture. Consequently two types of applications are appropriate for TYPELINE architecture. First is the applications that their datatype operations behave uniformly. Second target is applications that use approximate normal distribution of datatype operations. These application almost have user define datatypes and data structures. The best result in this part is referred to raytracer by approximately 28% and finally the worst result is related to btree with 11%. The here we can discuss about these result. In raytracer code regions, datatype operations have better distribution status in comparison with others. Btree uses only two of SDTs and should be appropriate for execution in TYPELINE. But when we have a look into the code regions, it is clear that execution codes according to datatype categories have a bad distribution and is not compatible with parallelism concepts of TYPELINE. The average parallelism percent is about 18%.

By parallelization that occur in various benchmarks, the reduction of execution cycles is computed approximately. The average percentage of cycle reduction is about 20%. It is very significant to know reasons of cycle reduction. In application-specific environments like raytracer benchmark that datatypes are distributed in a better way, all load and computational instructions are more effective in interaction to the process lines. The overall cycle reduction is led to enhance entire speed of execution.

The last point in this section is to review traditional execution out of TYPELINE. In the mentioned benchmarks, percentage of miss-handling is varied from 2% to 5%. This datatype operations include pointers and special type definitions that are not compatible with process lines that should handle in traditional process line.

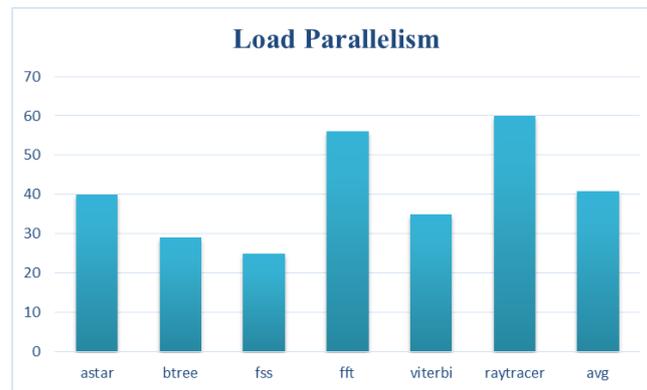

(a)

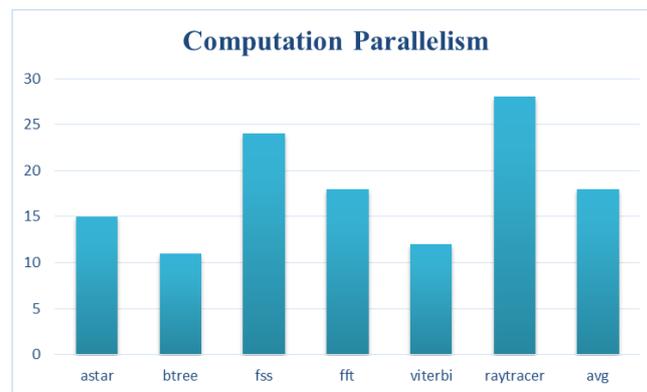

(b)

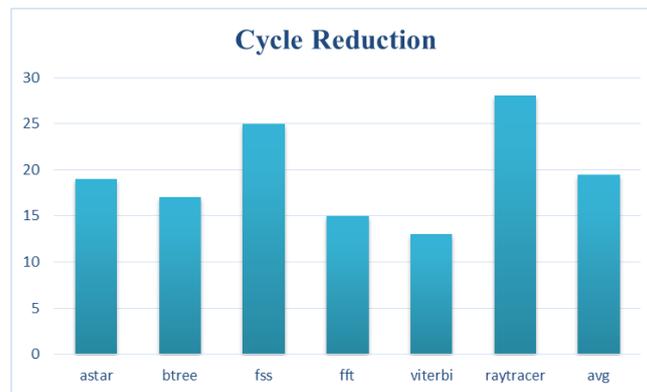

(c)

Figure 6: Experimental result after compilation process. (a) It shows load instruction parallelism under various benchmarks. The average parallelization is about 40% (b) Indicates instruction parallelism in computational area. Instructions include arithmetic and logical operation is given here. (c) Estimated cycle reduction due to parallelization and also datatype operation clustering.

## 5. Comparison and Analysis

In this section we compare TYPELINE with other prior works in this area. So it is shown that application-specific target like DNS servers are most suitable for implementation in process lines so the overall performance depends on this selection. Previous works like conservation core and Truffle in energy-efficient category, Parkour and Kremlin in parallelism and OOMIPS in are the main targets to compare with current architecture.

### 5.1. Parallelism

In this area we face to two categories in previous efforts. First is software optimization to know what regions in codes are deserved to parallelism and should be able to parallel execution. The Kremlin is work similar this [12]. The framework include discovery and planning before execution is applied by Kremlin tool. Instead Parkour is work on parallelization implementation on AMD 32-cores and it emphasized on performance [11].

TYPELINE parallelism procedure is focused on both software and hardware level. Notice that in comparison with Parkour that is tested by 32 and 64 parallel cores, TYPELINE is only implemented in four parallel process lines. In code optimization level it uses datatype operation clusters in two architecture support that is divided into load instruction clustering and computational instruction clustering. In average load instruction parallelism is about 18% and computational instruction parallelism is up to 40%. TYPELINE performance in parallelism is directly depends on datatype operation distribution. In environments like application specific programs, the overall parallelism improves.

### 5.2. Speed-up and Energy-efficiency

Accelerators are provided to speedup code execution. They reduce overall execution time with respect to architecture support in target applications. According to the parallelism results of TYPELINE and cycle reduction, it reduces overall cycles needed to execution about 20%. It leads to reduce CPI and improves speed of execution in general applications. It is important to know which code region and which datatype operation improve execution speed.

Another aspect is energy-efficiency. Some prior works in this area include Truffle and conservation core [1, 2]. C-core is an energy-efficient architecture that is synthesis-enabled and can be configured or patch in new situations. But TYPELINE have architecture support by using instruction set for each process line. This feature improves reliability of execution. First advantage is to have more flexibility in various types of codes and the second is better performance in process lines that only a few part of code regions cannot be supported in precise mode. This amount is changed from 2% up to 5%. Also Truffle is used low-power technique that works in two precise and approximate mode. This feature also can be implemented in TYPELINE. We can assign specific operational voltage to each process line. This comes from datatype features that are needed to handle the same operation groups. Using this technique can be one of the future works in this area to detect optimized the voltage of datatype operation.

### 5.3. TYPELINE and Dark Silicon Problem

The dark silicon constraints are effected on hardware/software design. The main problem is related to power that is needed to charge all silicones and also thermal management in processors [20, 24]. As mentioned before, when design space grows, we should ensure that all parallel units can be used with the same power resource. For example in 90 nm TSCM we can only use about 17% of silicones in the same time. This is optimization problem. In TYPELINE architecture, it is ensured that parallel execution units cannot exceed this amount. Although this should not be considered as a solution for Dark Silicon problem, but still this can be a prevention to use the entire available facilities. Also in future works that can be focused on energy-efficiency, TYPELINE will be a solution for Dark Silicon problem.

### 5.4. Application- Specific Review

In some aspects of modern industrial, computation-intensive applications and pervasive computing, datatype architecture can be a good target. These environments are almost application-specific. This feature is best for TYPELINE architecture implementation. As mentioned before, DNS servers and data center applications are most common targets of TYPELINE especially on energy-efficiency and Green computing area [25]. We know that every day billions of queries are submitted to the DNS servers and they response to them as a routine way. DNS datatype operations are uniform and can be modeled so easily. Another is modern vehicles that need more support on data acquisition and related area [26]. A very important area that is completely compatible to the TYPELINE and datatype theory is pervasive computing [27]. The idea behind this, is to propagate datatype operations among processors especially through CAN networks. Consider that we have some processors with various functionalities according to the target applications. Each processor that support operations in its own area, can also support other processors by its own datatype operations. So in this network, datatypes that are most common can be supported by maximum number of processors and the speed of computing will be enhanced due to datatype architecture design in uniform environments.

## 6. Conclusion

As operation datatype features in standard benchmarks, SDTs are extracted for implementation step. For each datatype operation have considered special process line. These process lines were responsible for computing operations that are supported by the same datatype. TYPELINE architecture is constructed with some process lines, register files related to each process line and finally TEUs that are responsible for executing instructions. TEUs are designed due to datatype features. In this procedure, more design space is dedicated to common datatypes and consequently more instructions are dedicated in the proposed ISA. Experimental results show that average parallelism can be improved in load instructions to 40% and in computation instructions up to 18%. Also it caused overall cycle reduction to the 20%.

It is shown that uniform and application-specific environments are the best targets for TYPELINE. These types of applications can improve overall performance of the framework with respect to better datatype operation distribution. Consequently it provides architecture support for complex data types that are constructed of the common datatypes. Structures, user-defined datatypes and arrays are the main supported features in this area. Also it shows that code dependencies can be more easily managed and clustered by datatype operations in both static and dynamic steps.

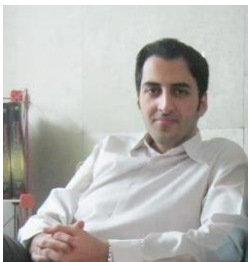

**Mehran Alidoost Nia** was born in Rasht, Iran in 1989. He studied computer software engineering at the University of Guilan in Rasht, Iran. He got his B.Sc degree in 2012 and at the same year he was accepted to follow his masters' study under exceptional talent office at Guilan University. His Msc project is about energy-efficient computer design using datatype modeling. His research interests include Energy-efficient Architectures, Advanced Datatype, Cloud Computing, Green Computing and Computer Security.

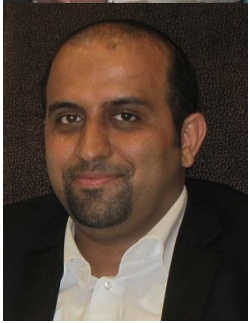

**Reza Ebrahimi Atani** was born in Arak, Iran. He studied Electronics Engineering at the University of Guilan in Rasht, Iran. He got his B.Sc degree in 2002 and at the same year he was accepted to follow his masters study at Iran University of Science & Technology (IUST) in Tehran, Iran. He joined the Electronics Research Center of IUST and received the M.Sc degree in Electronics-VLSI design in 2004. He received the Ph.D. degree in 2010 where he worked on "Design and Implementation of a stream cipher for mobile communications" as his PhD dissertation. He has an assistant professor position in Department of Computer Engineering at the University of Guilan. His research interests include Cryptology, Cryptographic Hardware and Embedded Systems (CHES), Computer and Network security, Information Hiding.